# Ngram-LSTM Open Rate Prediction Model (NLORP) and Error_accuracy@C metric:
## Simple effective, and easy to implement approach to predict open rates for marketing email

*Shubham Joshi*
*Principal Data Scientist, Oracle*
Email: shubham.s.joshi@oracle.com

*Indradumna Banerjee*
*Principal Data Scientist, Oracle*
Email: indradumna.banerjee@oracle.com

**Abstract**—Our generation has seen an exponential increase in digital tools adoption. One of the unique areas where digital tools have made an exponential foray is in the sphere of digital marketing, where goods and services have been extensively promoted through the use of digital advertisements. Following this growth, multiple companies have leveraged multiple apps and channels to display their brand identities to a significantly larger user base. This has resulted in products, worth billions of dollars to be sold online. Emails and push notifications have become critical channels to publish advertisement content, to proactively engage with their contacts. Several marketing tools provide a user interface for marketers to design Email and Push messages for digital marketing campaigns. Marketers are also given a predicted open rate for the entered subject line. For enabling marketers generate targeted subject lines, multiple machine learning techniques have been used in the recent past. In particular, deep learning techniques that have established good effectiveness and efficiency. However, these techniques require a sizable amount of labelled training data in order to get good results. The creation of such datasets, particularly those with subject lines that have a specific theme, is a challenging and time-consuming task. In this paper, we propose a novel Ngram and LSTM-based modeling approach (NLORPM) to predict open rates of entered subject lines that is easier to implement, has low prediction latency, and performs extremely well for sparse data. To assess the performance of this model, we also devise a new metric called '*Error_accuracy@C*' which is simple to grasp and fully comprehensible to marketers.

**Keywords- E-mail marketing, Subject line, Digital Marketing, Ngram , LSTM, NLP**

## I. Introduction

Emails and push messages are important digital marketing channels to sell products. Users generally first open the email message and then click the links present in the email content to purchase. Therefore, the subject line of the email is crucial in capturing the user's attention. Each email is sent to several users as part of a campaign, and the ratio of the number of times an email is opened to the total number of emails, is known as the open rate.[1-3] Predicting subject line open rates is one of the important features of any marketing tools.[4] The sender and the subject line are two crucial pieces of information that customers have about the email at first sight, i.e., before opening as in the inbox view. Therefore, these two criteria have a significant impact on whether a consumer chooses to open the email. Improving open rates is essential to business success because opening the email is a must before making a customer transaction.[5,6]

Marketers are constantly looking for compelling words for their subject lines [7-9] that will entice email recipients to open the email. The marketer wants to evaluate the performance of a new subject line after creating it. Thus, a model that predicts the open rates of the input subject line is extremely valuable to the marketers. [10,11] Along with open rate, a marketer also needs to know which phrases in the subject line make a difference. By examining these phrases, the marketer can determine which phrases result in higher open rates and which result in lower open rates and can then further refine the subject line accordingly to improve the open rates.

This paper focuses on designing a model to predict the open rates of an email by using the subject lines. Marketers can gain vital information about the effectiveness of the email campaign from the predicted open rates. Our algorithm not only produces the open rates but also highlights the relevant 3-word phrases (trigrams) and their open rates, explaining the predicted open rate for the supplied subject line. As an example, the subject line shown in Figure1, *'Last chance as the subject line Excellent summer getaways. Save up to 25%'*, our

$$\underbrace{\text{Last chance}}_{x1\,:\,17\%} - \underbrace{\textit{Great summer escapes}}_{x2\,:\,13\%}. \underbrace{\textit{Save up to 25\%}}_{x3\,:\,18\%}$$

$$x\ (open\ rate\ for\ SL) = \frac{(17\% + 13\% + 18\%)}{3}$$

*Figure 1 – Illustration of the process of subject line open rate calculation for the trigrams in an email subject line.*

model will predict an open rate $x$ as the average of $x1, x2$ and $x3$ open rates, where $x1, x2$ and $x3$ are defined as the open rates of important non-overlapping 3 words phrases in the subject line. Developing a model to predict open rates using just subject lines is a big challenge, especially if the data is sparse and the subject lines are non-repeating. Some machine learning algorithms that have already been tested for a similar domain (ad click prediction), includes the support vector machine (SVM), logistic regression and decision trees [12,13]. Factorization machines have been used to capture non-linear relationship amongst features for adclick predictions.[14] Deep learning neural networks such as Convolutional neural networks have been used to detect complex interaction between features gathered from user behaviours. [15,16] Creating a model to predict open rates using only subject lines is difficult [17,18] since the model must learn trends from the few words that are included in the subject lines in order to forecast the open rate, which should be accurate even with unknown words. In this paper, we present a combined Ngram and LSTM based approach to predict open rates for subject lines which is simple, effective for sparse data, supports multiple languages and is easy to implement. However, merely having an elegant model does not suffice, as we also need to have a metric which explains the performance of the model and is intuitive and easy to understand for the marketers. Therefore, we also propose a new metric Error_accuracy@C to measure the performance of open rate prediction model, which is easier to interpret and explain to the marketers. Here 'C' represents the error tolerance limit, which can be decided by the business or user using it.

## II. Methods
### A. General ideas and notions

Our prediction of open rates from subject lines is based on the historical performance of phrases present in the subject line. In this method, we consider phrases of a single word (unigram), two words (bigram), and three words (trigram). The phrases with high open rates in historical data are likely to have higher open rates in the future. We average the open rates of the top five non-overlapping trigrams from historical data to get the subject line's final open rates. The open rate for each of these trigrams is calculated as the average of the trigram's historical open rates, and historical open rates of all bigrams and unigrams present in the trigram.

The historical open rates of the phrases are obtained by taking the average of the open rates of the subject lines in which the phrase has occurred in historical data and save them as a mapping file - $M_f$. We will call the subject line for which we need to predict the open rates as *input subject line*. The mapping file ($M_f$) is used during prediction to retrieve the historical open rates of the phrases present in the *input subject line*. Let's consider $T_n$, $B_n$ and $U_n$ be a set of trigrams, bigrams and unigrams present in the subject line (collectively referred as $P_n$) for which we need to predict the open rates. If we consider $T_1, B_1$ and $U_1$ as some of the phrases than the open rates of the phrases denoted by $X_{t1}, X_{b1}$ and $X_{u1}$ are obtained as,

$$X_{t1} = M_f(T_1)$$
$$X_{b1} = M_f(B_1)$$
$$X_{u1} = M_f(U_1)$$

We will be following the same notations throughout this paper.

Now what happens when mapping for a phrase is not present in the mapping file $M_f$? These phrases are tagged as *out of vocabulary phrases* (OOP). To obtain the open rates for out of vocabulary phrases, we train a LSTM based deep learning regression model using phrases from $M_f$ as features and corresponding open rates as label. Let's call this model as $LSTM_{orp}$, which is an LSTM based model to predict open rates of phrases. Thus, for an out of vocabulary phrase present in the *input subject line* we will predict its open rate using this LSTM model. For example, let's say $T_1$ is an OOP then we will use the logic mentioned below, and the *else* condition in the logic will be executed as shown in algorithm 1 below.

| Algorithm 1: Open rate calculation logic |
|---|
| Input: $M_f(T_1)$ and $LSTM_{orp}(T_1)$ |
| if present $M_f(T_1)$ then<br>   return $M_f(T_1)$<br>else<br>   return $LSTM_{orp}(T_1)$<br>end |

Thus, the open rate for a trigram $X_{tn}$ is obtained as,

$$X_{tn} = Avg(M_f(T_n))$$
$$+ Avg(\sum_{m=0}^{m=M} \begin{cases} M_f(B_{tn,m}) \text{ if preset} \\ LSTM_{orp}(B_{tn,m}) \end{cases})$$
$$+ Avg(\sum_{l=0}^{l=L} \begin{cases} M_f(U_{tn,l}) \text{ if preset} \\ LSTM_{orp}(U_{tn,l}) \end{cases}))$$

After obtaining open rates for all the trigrams we sort the trigrams by open rate and pick top 5 non-overlapping and take average of the open rates of these trigrams to get over all open rates.

$$X_{sorted} = sort([x_{t1}, x_{t2}, x_{t3} \ldots \ldots x_{tn}])$$
$$X_{no-overlap} = remove\_overlapping(X_{sorted})$$
$$X_{final} = \sum_{k=1}^{k=5} x_{no-overlap,k}$$

B. *Error_Accuracy@C and Average percentage Error*

Error (E) is defined as the absolute difference between the actual and predicted values of open rates.

$$E = |actual\ open\ rats - predicted\ open\ rate|$$

This error term is calculated for each subject line present in the test set. These error terms represent a verry crucial information – how much the model predictions are deviating from the actual open rates. Each marketer may have a different tolerance limit for the error which is based on the product domain. We call this as confidence cutoff (C).

The percentage of predictions which lie within the confidence cutoff is defined as Error_accuracy@C. This tells - the percentage of predictions within acceptable limit of error.

$$Error\_accuracy@C = \frac{predictions < C}{total\ predictions}$$

Along with this the average % error which is defined as :-

$$Average\ \%\ Error = mean(\frac{|Actual\ open\ rate - Predicted\ opem\ rate|}{Actual\ open\ rate})$$

Tells percentage of overall error.

C. *Training and Prediction flow*

The training flow consists of extracting unigram, bigram, and trigrams from each subject line from the training set and saving it along with the subject line in which they occur and the open rates of the subject line in the following format: phrase, subject_line, open_rate. This is followed by filtering out stop words from unigram. Next we calculate average open rates for each phrase (unigram, bigram or trigram), and store this as a mapping with key as phrase and value as average open_rate for that phrase.

The next step is training a 3-layer LSTM model (figure 2) model to predict open rates for the phrase, which will be used when any out of vocabulary token is found during prediction. The entire training flow is depicted in figure 3, showing the data extraction, phrase generation, model training and file saving steps of the workflow. The model outputs a mapping file comprising of the phrases and their corresponding open rates, and outputs the *LSTMorp* model.

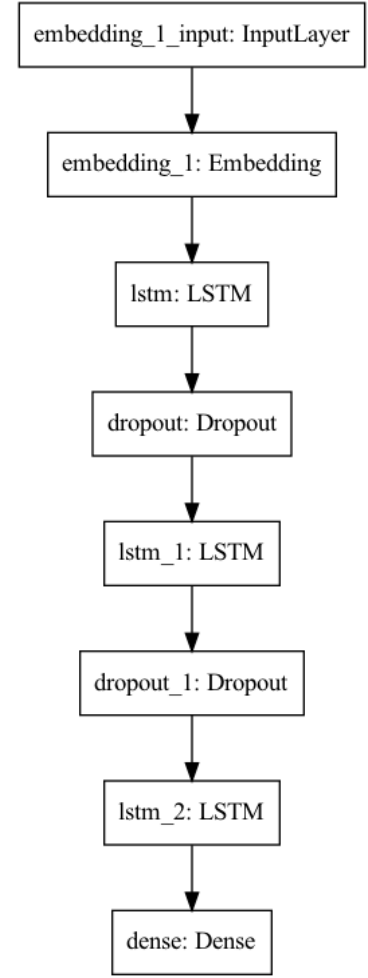

*Figure 2* – *A three Layer LSTM Model for predicting open rates for phrases occurring in subject lines, comprising of LSTM, embedding and dropout layers*

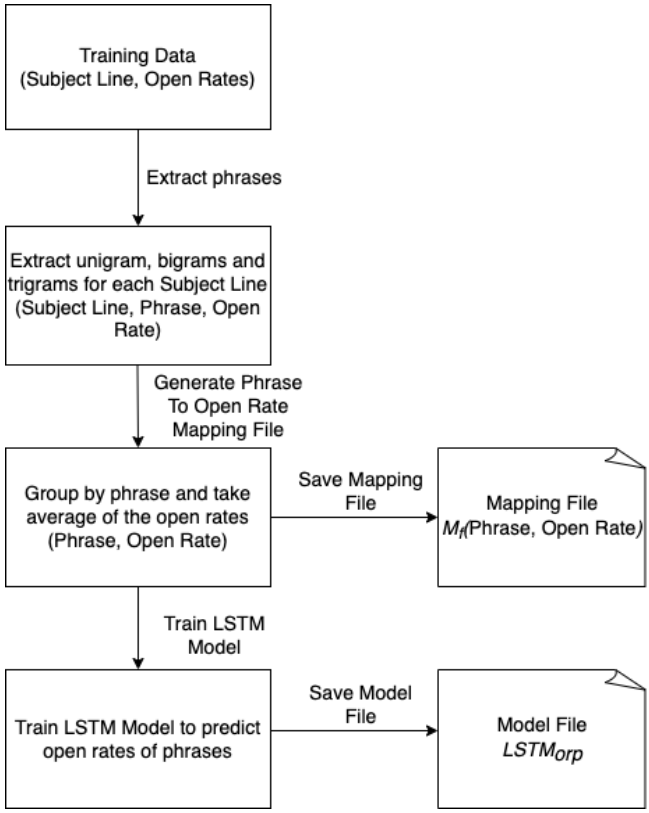

*Figure 3* – *Flowchart representing the training flow of the NLORP model that starts with the phrase extraction, and gives out a mapped file and the LSTMorp model.*

The training flow is followed by the prediction flow of the model (Figure 4). The prediction flow involves extracting phrases (unigram bigram and trigram) from input subject line. For each phrase (unigram, bigram and trigram), we get the open rate from the mapping file $M_f$, that was generated in figure 3 above.

We follow the same logic described in Algorithm 1 earlier for choosing the right model and then calculate the open rate for each trigram using the following steps.

i. If phrase is not present in the mapping, predict the open rates using LSTM$_{orp}$ model.

ii. Calculate open rate for each trigram by taking average of the open rate of trigram as well as the unigrams and bigrams present in the trigram.

iii. Take top 5 non-overlapping trigrams and take average of the open rates of these trigrams to get the overall open rate.

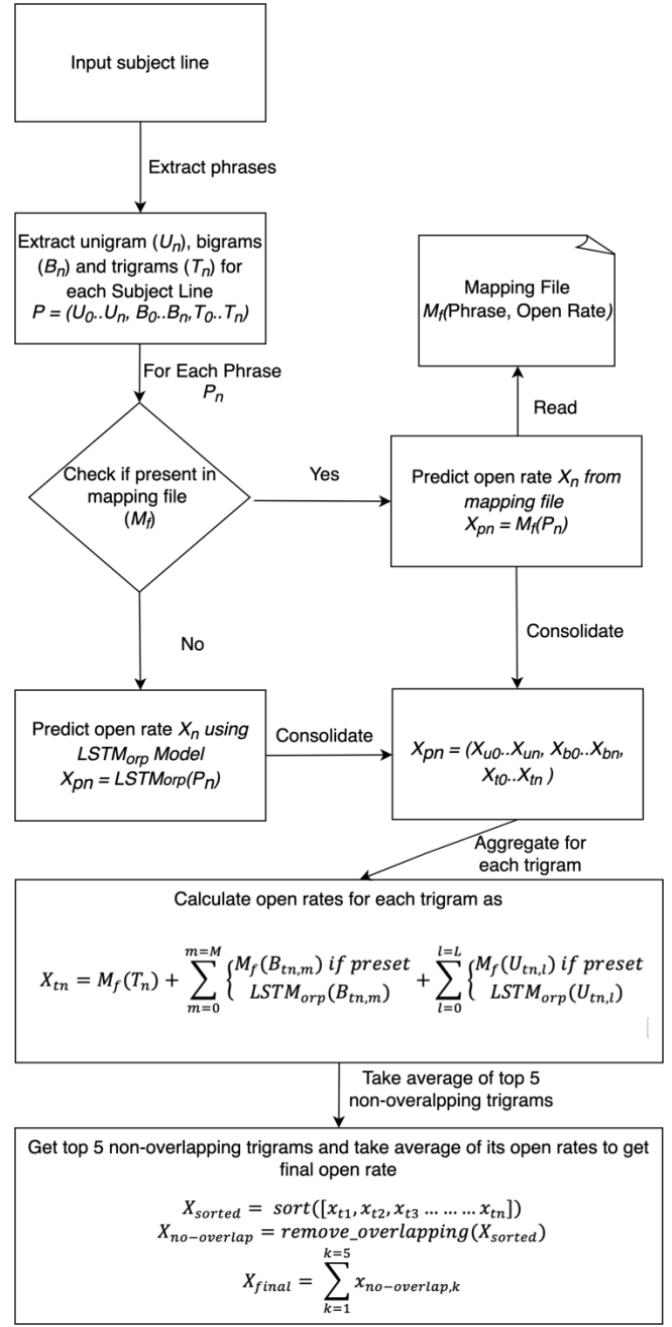

*Figure 4* – *Prediction Flow which starts with extracting the subject line phrases, predicting open rate from Mapping files or LSTMorp model, consolidating aggregates for each trigram and averaging them.*

### D. Dataset and Data Preprocessing

The email subject line dataset comprises of 300+ different subject lines of special deal emails, that were picked up from multiple internet sources via google search. The dataset was chosen in a way keeping in mind the general audience, to whom a variety of deals have been offered, and the deals can be on wide ranging topics that include discounts on cars, exclusive dress deals, earning shopping points, movie related information and deals, and email marketing software related information to name a few. The typical character length for these subject lines was anywhere between 15 to 60 characters, and all the subject lines were in English language. For a better improved version, of the model, in later iterations we would also like to include other languages, specially languages with non-English script such as Japanese and Chinese. This will be an interesting study because the constructs and grammar used in English are typically not used in Japanese, Chinese, and many other scripts, making extracting phrases from such scripts a different challenge. However, in this paper the scope of LSTM$_{orp}$ model was strictly restricted to English language subject lines.

III. **Results and Discussion**

A. *Model Performance Visualization*

The performance of this model is measured in terms of Error_accuracy@C, and average percentage error that has been already described above. Here C is the tolerance cutoff, were,

$Error = |actual\_open\_rate - predicted\_open\_rate|$

The detailed explanation of the whole process to evaluate the model performance is described here. Firstly, after getting the predictions we calculate the error as mentioned above. Next we decide the value of confidence cut off - C. For this work as an example, taking C=0.1 shows a tolerance of 10% Error. Next, we create 2 group of our data, Group1: Having Error<=0.1 and Group2: Having Error>0.1. The metrics for each group are then calculated using Error Accuracy @C percentage of subject lines in each group. The average error percentage, which is mean of absolute value of the difference of the actual Open Rates and Predicted Open Rates to the actual Open Rates. In an ideal scenario, the error accuracy should be maximum for 'Error<=0.05' group, and Average % error should be as less as possible.

The model is trained using 5 fold cross validation, with performance metrics calculated on each of the 5 test folds. The average of these metrics is used to calculate overall performance metrics, plotted as shown in Figure 5 below.

Plot shown in Figure 4 is used to visualize the performance of the model. Here the left pie chart is used to visualize Error_accuracies@C metric. Since our final open rate is calculated as the average of the open rate of top 5 non-overlapping phrases as described above. The Green Section

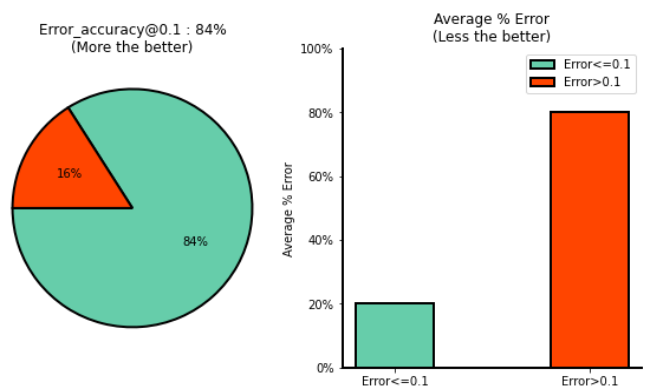

*Figure 5* – *Model Performance Visualization of the Group 1 and Group 2 error accuracies vs the average percentage error for each of the two groups (Group1: Error<0.1, and Group2: Error>0.2)*

represents the percentage of the count of subject lines for which the Error >= 0.1 (84%), whereas Red Section represents the percentage of subject lines with Error < 0.1 which is 16%. The bar plot at the right is visualize the Average % error. For all subject lines with Error<=0.1 the Average % error is 20%, whereas for all subject lines with Error>0.1 the Average % error is 80%. This performance metric clearly demonstrates that our model is performing extremely well.

B. *Prediction Explainability*

We can use the same approach, as earlier shown in figure 1, to explain our prediction as:

$$x \ (open \ rate \ for \ SL) = \frac{(17\% + 13\% + 18\%)}{3}$$

*Figure 6* - *Subject line open rate calculation for final open rate calculation for trigrams in an email subject line.*

Thus, for each subject line, the open rates of the top five important three-word phrases can be displayed to the marketer on the front end where this model will be consumed, providing the marketer with a good understanding of the effectiveness of the specific phrase to be used in the subject line of the marketing email.

## IV. CONCLUSION

The generation of relevant subject lines, that is appealing to consumers is an integral component of the multi-billion email marketing industry. An option to predict the email subject line open rate will help marketers in sending more user friendly and acceptable subject lines, in turn leading to better revenues for any company. In this paper, we present a novel combined Ngram and LSTM based approach that is very useful for real life scenarios where email subject lines are not very frequently repeated, or are sparse in nature. We also present a robust metric Error_accuracy@C to explain the performance of our model, and that is intuitive and easy to understand for the marketers.

The future directions that this work can take includes solving related problems and targeting personalized email subject lines specific to users. This will mean that the marketer will not be sending the same email subject to every consumer, but rather customise the subject lines, based on what phrase in the email content is appealing to the customer. The methods described in our work here may be easily extended to personalised email campaigns.

## V. REFERENCES

1. Paulo, M., Vera L. Miguéis, and Ivo Pereira. "Leveraging email marketing: Using the subject line to anticipate the open rate." *Expert Systems with Applications* 207 (2022): 117974.
2. Teiu, Codrin. "Email subject lines analysis for high open rate in email marketing." New Trends in Sustainable Business and Consumption (2020): 835.
3. Sappleton, Natalie, and Fernando Lourenço. "Email subject lines and response rates to invitations to participate in a web survey and a face-to-face interview: the sound of silence." International Journal of Social Research Methodology 19, no. 5 (2016): 611-622.
4. Balakrishnan, Raju, and Rajesh Parekh. "Learning to predict subject-line opens for large-scale email marketing." In *2014 IEEE International Conference on Big Data (Big Data)*, pp. 579-584. IEEE, 2014.
5. Bala, Madhu, and Deepak Verma. "A critical review of digital marketing." *M. Bala, D. Verma (2018). A Critical Review of Digital Marketing. International Journal of Management, IT & Engineering* 8, no. 10 (2018): 321-339.
6. Conceição, Andreia, and João Gama. "Main factors driving the open rate of email marketing campaigns." In *International conference on discovery science*, pp. 145-154. Springer, Cham, 2019.
7. Scholz, Michael, Joachim Schnurbus, Harry Haupt, Verena Dorner, Andrea Landherr, and Florian Probst. "Dynamic effects of user-and marketer-generated content on consumer purchase behavior: Modeling the hierarchical structure of social media websites." *Decision Support Systems* 113 (2018): 43-55.
8. Aiossa, Ewelina. "The anatomy of an effective e-mail subject line: How to stand out in a crowded inbox." *Journal of Digital & Social Media Marketing* 8, no. 3 (2020): 244-250.
9. Obermiller, Carl. "Marketing Benchmarks: Do You Trust Your Friendly Marketer?." *Journal of Consumer Affairs* 53, no. 1 (2019): 71-86.
10. Zhang, Rui, and Joel Tetreault. "This email could save your life: Introducing the task of email subject line generation." *arXiv preprint arXiv:1906.03497* (2019).
11. Araújo, Carolina, Christophe Soares, Ivo Pereira, Duarte Coelho, Miguel Ângelo Rebelo, and Ana Madureira. "A Novel Approach for Send Time Prediction on Email Marketing." Applied Sciences 12, no. 16 (2022): 8310.
12. Riana, Dwiza. "Deep Neural Network for Click-Through Rate Prediction." *International Journal of Software Engineering and Computer Systems* 8, no. 2 (2022): 33-42.
13. McMahan, H. Brendan, Gary Holt, David Sculley, Michael Young, Dietmar Ebner, Julian Grady, Lan Nie et al. "Ad click prediction: a view from the trenches." In *Proceedings of the 19th ACM SIGKDD international conference on Knowledge discovery and data mining*, pp. 1222-1230. 2013.
14. Rendle, Steffen. "Factorization machines." In *2010 IEEE International conference on data mining*, pp. 995-1000. IEEE, 2010.
15. Liu, Qiang, Feng Yu, Shu Wu, and Liang Wang. "A convolutional click prediction model." In *Proceedings of the 24th ACM international on conference on information and knowledge management*, pp. 1743-1746. 2015.
16. Liu, Bin, Ruiming Tang, Yingzhi Chen, Jinkai Yu, Huifeng Guo, and Yuzhou Zhang. "Feature generation by convolutional neural network for click-through rate prediction." In *The World Wide Web Conference*, pp. 1119-1129. 2019.
17. Jaidka, Kokil, Tanya Goyal, and Niyati Chhaya. "Predicting email and article clickthroughs with domain-adaptive language models." In Proceedings of the 10th ACM Conference on Web Science, pp. 177-184. 2018.
18. Abakouy, Redouan, El Mokhtar En-Naimi, Anass El Haddadi, and Lotfi Elaachak. "Machine Learning as an Efficient Tool to Support Marketing Decision-Making." In The Proceedings of the Third International Conference on Smart City Applications, pp. 244-258. Springer, Cham, 2020.